\documentclass[atmp]{ipart_v1}

\Vol{19}
\Issue{3}
\Year{2015}
\firstpage{701}

\usepackage{t1enc}
\usepackage[latin1]{inputenc}
\usepackage[english]{babel}
\usepackage{epstopdf}

\usepackage{amsthm}
\usepackage{yfonts}

\usepackage{bbm}
\usepackage{bm}
\usepackage{mathrsfs}

\newcommand{\be}[0]{\begin{equation}}
\newcommand{\ee}[0]{\end{equation}}

\numberwithin{equation}{section}

\theoremstyle{plain}

\begin{document}

\title[Local fractional Klein-Gordon equation]{Analytical solution of local fractional Klein-Gordon equation for the generalized Hulthen potential}

\author[H. Karayer \textit{et al}]{Hale Karayer, Dogan Demirhan and Fevzi Buyukkilic}

\begin{abstract}
One dimensional Klein-Gordon (KG) equation is investigated in the domain of conformable fractional calculus for one dimensional scalar potential namely generalized Hulthen potential. The conformable fractional calculus is based on conformable fractional derivative which is the most natural definition in non integer order  calculus. Fractional order differential equations can be solved analytically by means of this derivative operator. We obtained exact eigenvalue and eigenfunction solutions of local fractional KG equation and investigated the evolution of relativistic effects in correspondence with the fractional order.
\end{abstract}

\maketitle

\section{Introduction}
\label{intro}
Relativistic wave equations namely  Dirac equation and  KG equation have great importance in the efforts to determine dynamics of a relativistic particle in relativistic quantum mechanics. 
Solution of the KG equation explains the behavior of a spinless particle of rest mass $m$  at high energies and velocities comparable to the speed of light. Bound state solutions of the KG equation have been studied by many authors in the literature, see \cite{yasuk,egrifes,diao,akbarieh,motavalli,antia} and references therein. Different methods can be used to obtain exact or approximate solutions of KG equations which are written for various potential functions. The Nikiforov-Uvarov (NU) method, Supersymmetric Quantum Mechanics, Factorization Method and Asymptotic Iteration Method are the most frequently used methods \cite{ikot}.

Using the theory of fractional calculus  which is based on non integer order differentiations and integrations, a lot of physical phenomena can be described successfully \cite{ganji,tenreiro,sier,podlubny}. Consequently differential equations which describe physical systems are handled in the fractional domain.  Various definitions have been proposed for the fractional order differential and integral operators. 
Predominant definitions are Riemann-Liouville and Caputo definitions \cite{miller}.
\begin{eqnarray}
\label{rl}
^{RL}D_{a}^{\mu}f(x)=\frac{1}{\Gamma(n-\mu)}\frac{d^{n}}{dx^{n}}\int_{a}^{x}(x-t)^{n-\mu -1}f(t)dt, \nonumber \\
\label{caputo}
^{C}D_{a}^{\mu}f(x)=\frac{1}{\Gamma(n-\mu)}\int_{a}^{x}(x-t)^{n-\mu -1}f^{(n)}(t)dt.\nonumber
\end{eqnarray}
where   $\mu \in R$, $n-1 \leq \mu < n$ and the superscripts $RL$ and $C$ stand for Riemann-Liouville and Caputo respectively. These are non local operators and do not satisfy classical properties such as chain, product and quotient rules which allow us to achieve analytical solution in the standard calculus. In 2014 a local form of fractional derivative operator is defined by Khalil et al. \cite{khalil}. This is the most natural fractional order derivative operator which provides the above mentioned rules. Thereafter conformable fractional calculus theory \cite{abdaljawad}, applicability of this definition in quantum mechanics \cite{anderson}, solution of fractional Schr\"odinger equation \cite{eslami} are studied in the view of this local fractional derivative definition. Though the definition fails some properties which are pointed out by Ortigueira and Machado \cite{ort}, it is more suitable for applications as compared with Riemann-Liouville or Caputo fractional derivative operators for real physical problems \cite{iyola}.

In a recent work, we have derived conformable fractional form of the NU method and solved local fractional  Schr\"odinger equation for harmonic oscillator potential, Hulthen potential and Woods-Saxon potential in order to present accuracy of the method \cite{karayer}.

The aim of the present work is to solve fractional order one dimensional time independent KG equation for the generalized Hulthen potential in the scalar coupling scheme using the conformable fractional NU method. The manuscript is organized as follows; In Sec.~\ref{sec:1} the formalism of KG equation for the generalized Hulthen potential is briefly outlined. In Sec.~\ref{sec:2} the definition of conformable fractional derivative operator and conformable fractional NU method are reviewed. In Sec.~\ref{sec:3} we have presented analytical solution of fractional KG equation for the generalized Hulthen potential. Finally, conclusions are discussed in the last section.
\section{Formalism of the KG equation with the generalized Hulthen potential in scalar scheme}
\label{sec:1}
One dimensional time independent KG equation for a spinless particle of rest mass $m$ in the presence of vector and scalar potentials is given by;
\begin{equation}
\psi''(x)+\frac{1}{\hbar^2 c^2}\bigg[\big(E-V(x)\big)^2 -\big(mc^2+S(x)\big)^2\bigg]\psi(x)=0,
\end{equation}
where $V(x)$ and $S(x)$ are vector and scalar potentials respectively. For the existence of bound state solutions it is required that $S(x)>V(x)$ \cite{egrifes}. When $V(x)=0$, the one dimensional KG equation for a given scalar potential $S(x)$ is reduced to the following form;
\begin{equation}\label{skg}
\psi''(x)+\frac{1}{\hbar^2 c^2}\bigg[E^2 -\big(mc^2+S(x)\big)^2\bigg]\psi(x)=0,
\end{equation}
In this case Eq.(\ref{skg}) can be transformed to a second order Schr\"odinger-like differential equation;
\begin{equation}\label{slkg}
\psi''(x)+\frac{2m}{\hbar^2 }\bigg[E_{eff} -U_{eff}(x) \bigg]\psi(x)=0,
\end{equation}
where $E_{eff}$ and $U_{eff}$ are effective energy and effective potential given by;
\begin{equation}
E_{eff}=\frac{E^2-m^2c^4}{2mc^2},\,\,\,\,\,\,U_{eff}(x)=\frac{S^2(x)}{2mc^2}+S(x).
\end{equation}
Therefore Eq.(\ref{skg}) can be rewritten in the following form for $\hbar=c=1$ \cite{egrifes};
\begin{equation}\label{slkg1}
\psi''(x)+\big[-S^2(x)-2mS(x)-(m^2-E^2)\big]\psi(x)=0.
\end{equation}
In order to specify dynamics of a relativistic particle in a scalar potential, the potential function $S(x)$ is inserted in this equation. Here the potential function is chosen as generalized Hulthen function which is given by;
\begin{equation}\label{ghp}
S(x)=-S_{0}\frac{e^{-\alpha x}}{1-qe^{-\alpha x}}
\end{equation}
where $q$ is a deformation parameter. This potential transforms to exponential potential, standard Hulthen potential and Woods-Saxon potential for $q=0$, $q=1$ and $q=-1$ respectively. Substituting the potential function given by Eq.(\ref{ghp}) in Eq.(\ref{slkg1}) and using a transformation  $z=S_{0}e^{-\alpha x}$  the following hypergeometric type differential equation is obtained \cite{egrifes};
\begin{eqnarray}\label{cfkghp}
\psi_{q}''(z)+\frac{S_{0}-qz}{z(S_{0}-qz)}\psi_{q}'(z)+
\frac{1}{\big[z(S_{0}-qz)\big]^{2}}\big[-(\gamma^2+\nonumber \\
q\beta^2+q^2\epsilon^2)z^2+S_{0}(\beta^2+2q\epsilon^2)z-S_{0}^2 \epsilon^2\big]\psi_{q}(z)=0
\end{eqnarray}
for which
\begin{equation}\label{qntts}
\gamma^{2}=\frac{S_{0}^{2}}{\alpha^{2}}, \,\,\,\,\,\,\beta^{2}=\frac{2mS_{0}}{\alpha^{2}},\,\,\,\,\, \epsilon^{2}=\frac{1}{\alpha^{2}}(m^{2}-E^{2}).
\end{equation}
\section{A brief review of the conformable fractional derivative operator and conformable fractional NU method}
Local fractional derivative operator which is a natural extension of the standard derivative definition, is introduced by  Khalil et al. for the first time;

\label{sec:2}
\begin{eqnarray}\label{eq1}
D^{\mu}[f(t)]&=& \lim\limits_{\epsilon \rightarrow 0}\frac{f(t+\epsilon t^{1-\mu})-f(t)}{\epsilon}, \,\,\,\,\,t>0\\
\label{rightlimit}
f^{(\mu)}(0)&=& \lim\limits_{t \rightarrow 0^{+}}f^{(\mu)}(t)
\end{eqnarray}
where $0<\mu \leq 1$ and $D^{\mu}$ is the local fractional derivative operator \cite{khalil}. This operator provides the basic rules such as product, quotient, chain rules which are valid in standard calculus;
\begin{eqnarray}
D^{\mu}[af+bg]=aD^{\mu}[f]+bD^{\mu}[g] & & linearity \nonumber \\
D^{\mu}[fg]=fD^{\mu}[g]+gD^{\mu}[f] & & product \,\, rule \nonumber \\
D^{\mu}[f(g)]=\frac{df}{dg}D^{\mu}[g] & & chain \,\, rule \nonumber \\
D^{\mu}[f]=t^{1-\mu}f' & & {where} \,\,\,\, f'=\frac{df}{dt}. \nonumber 
\end{eqnarray}
The last property is called as key property of the definition. If $f$ is differentiable, then $\mu th$ order derivative of $f$ is equal to product of its first order derivative with $t^{1-\mu}$. 
Since fractional order differential equations have great importance in describing physical systems with a realistic approach, some appropriate methods are derived to solve these equations. The NU method is a well known method which gives exact solutions of second order linear differential equations. In quantum mechanics the method has been used to solve Schr\"odinger-like differential equations for various potentials. The method is based on reducing the handled equation to a hypergeometric type second order differential equation;
\begin{equation}
\label{eq:nu}
\psi''(z)+\frac{\widetilde{\tau}(z)}{\sigma(z)}\psi'(z)+\frac{\widetilde{\sigma}(z)}{\sigma^{2}(z)}\psi(z)=0
\end{equation}
where $\widetilde{\tau}(z)$ is a polynomial of at most first-degree, $\sigma(z)$ and $\widetilde{\sigma}(z)$ are polynomials of at most second-degree and $\psi(z)$ is a function of hypergeometric-type \cite{Nikiforov}.
Then the reduced equation which is called as basic equation of the method, can be solved systematically by means of special orthogonal functions and eigenstate solutions can be achieved completely \cite{Nikiforov,1999a,egrifes b,ikhdair}.

Conformable fractional form of this method is introduced in order to solve conformable fractional order  Schr\"odinger equation and presented in our recent work \cite{karayer}. In the case of the conformable fractional NU method fractional orders are inserted in the basic equation. Then using the key property of the conformable fractional derivative operator one can reach to the following second order differential equation;
\begin{equation}
\label{eq:fnukd}
\psi''(z)+\frac{\widetilde{\tau}_{f}(z)}{\sigma_{f}(z)}\psi'(z)+\frac{\widetilde{\sigma}(z)}{\sigma_{f}^{2}(z)}\psi(z)=0.
\end{equation}
where $\widetilde{\tau}_{f}(z)=(1-\mu)z^{-\mu}\sigma(z)+\widetilde{\tau}(z)$ and $\sigma_{f}(z)=z^{1-\mu}\sigma(z)$ and the subscript $f$ stands for fractional. Boundary conditions of the conformable fractional NU method is determined by the degrees of the coefficients in the basic equation of the method given by Eq.(\ref{eq:fnukd}). Here $\widetilde{\tau}_{f}(z)$ is a function of at most  $\mu th$ degree  (which means that this function can also be equal to a constant), $\sigma_{f}(z)$ is a function of at most  $(\mu+1) th$ (i.e. the degree of this function can also be equal to $1$) and $\widetilde{\sigma}(z)$ is a function of at most $2\mu th$ degree (i.e. the degree of this function can also be equal to $0$ or $\mu$). If any fractional order differential equation is reduced to the basic equation using the key property of local fractional derivative operator, then it can be solved analytically by the conformable fractional NU method.

After determining the following newly defined functions related to the initial functions in the basic equation, eigenvalue and eigenfunction solution of Eq.(\ref{eq:fnukd}) can be obtained;
\begin{equation}
\label{eq:fpi}
\pi_{f}(z)=\frac{\sigma_{f}'(z)-\widetilde{\tau}_{f}(z)}{2}\pm\sqrt{(\frac{\sigma_{f}'(z)-\widetilde{\tau}_{f}(z)}{2})^{2}-\widetilde{\sigma}(z)+k(z)\sigma_{f}(z)}.
\end{equation}
Recall that $\pi_{f}(z)$ is  a function of at most  $\mu th$  degree. Providing this condition the expression under the square root sign must be the square of a $\mu th$ order function. Thus the function $k(z)$ under the square root sign must be chosen properly.
\begin{equation}
\label{eq:fh}
\tau_{f}(z)=\widetilde{\tau}_{f}(z)+2\pi_{f}(z).
\end{equation}
\begin{equation}
\label{eq:fk}
\lambda(z)=k(z)+\pi_{f}'(z).
\end{equation}
\begin{equation}
\label{eq:fp}
\lambda_{n}(z)=-n\tau_{f}'(z)-\frac{n(n-1)}{2}\sigma_{f}''(z)   \,\,\,\,\,\,\,\,   (n=0,1,2,...).
\end{equation}
In order to obtain eigenvalue solution, the function  $ \lambda(z) $ in Eq.(\ref{eq:fk}) is taken equal to $ \lambda_{n}(z) $ in Eq.(\ref{eq:fp}). For the eigenfunction solution, functions $\phi(z)$ and $y_{n}(z)$ given by;
\begin{equation}
\label{fi}
\frac{\phi'(z)}{\phi(z)}=\frac{\pi_{f}(z)}{\sigma_{f}(z)},
\end{equation}		
\begin{equation}
\label{rho}
(\sigma_{f}(z)\rho(z))'=\tau_{f}(z)\rho(z).
\end{equation}
\begin{equation}
\label{rod}
y_{n}(z)=\frac{B_{n}}{\rho(z)}\frac{d^{n}}{dz^{n}}[\sigma_{f}^{n}(z)\rho(z)],
\end{equation}
are inserted in $\psi(z)=\phi(z)y(z)$. 
\section{Solution of conformable fractional KG equation for the generalized Hulthen potential}
\label{sec:3}
Conformable fractional form of one dimensional KG equation for the generalized Hulthen potential given by Eq.(\ref{cfkghp}) is written by replacing integer orders with fractional orders;
\begin{eqnarray}\label{cfkghp1}
D^{\mu}D^{\mu}\psi_{q}(z)+\frac{S_{0}-qz^{\mu}}{z^{\mu}(S_{0}-qz^{\mu})}D^{\mu}\psi_{q}(z)+
\frac{1}{\big[z^{\mu}(S_{0}-qz^{\mu})
\big]^{2}}\big[-(\gamma^2+\nonumber \\
q\beta^2+q^2\epsilon^2)z^{2\mu}+S_{0}(\beta^2+2q\epsilon^2)z^{\mu}-S_{0}^2 \epsilon^2\big]\psi_{q}(z)=0
\end{eqnarray}
Using the key property of the conformable fractional derivative definition, Eq.(\ref{cfkghp1}) can be transformed to a second order differential equation;
\begin{eqnarray}\label{cfkghp2}
\psi_{q}''(z)+\frac{(S_{0}-qz^{\mu})(2-\mu)}{z(S_{0}-qz^{\mu})}\psi_{q}'(z)+
\frac{1}{\big[z(S_{0}-qz^{\mu})\big]^{2}}\big[-(\gamma^2
+q\beta^2+ \nonumber \\
q^2\epsilon^2)z^{2\mu}+S_{0}(\beta^2+2q\epsilon^2)z^{\mu}-S_{0}^2 \epsilon^2\big]\psi_{q}(z)=0.
\end{eqnarray}
Comparing this equation with the basic equation of the method, the parameters in Eq.(\ref{eq:fnukd}) are determined as;
\begin{eqnarray}\label{prmtr}
\widetilde{\tau}_{f}(z)=(S_{0}-qz^{\mu})(2-\mu) \nonumber \\
\sigma_{f}(z)=z(S_{0}-qz^{\mu})\nonumber \\
\widetilde{\sigma}(z)=-(\gamma^2+q\beta^2+q^2\epsilon^2)z^{2\mu}+S_{0}(\beta^2+2q\epsilon^2)z^{\mu}-S_{0}^2 \epsilon^2. 
\end{eqnarray}
Since $\widetilde{\tau}_{f}(z)$, $\sigma_{f}(z)$ and $\widetilde{\sigma}(z)$ are $\mu th$, $(\mu +1)th$ and $2\mu th$ order, conformable fractional NU method can be used in order to obtain the bound state solutions of local fractional KG equation for the generalized Hulthen potential. After substituting the parameters given by  Eq.(\ref{prmtr}) into Eq.(\ref{eq:fpi}), the function $\pi_{f}(z)$ can be obtained as;
\begin{eqnarray}
\label{eq:fpihp}
\pi_{f}(z)=\frac{1}{2}\big\{(\mu -1)S_{0}-qz^{\mu}(2\mu-1)\pm \nonumber \\
\big[[q^{2}(2\mu-1)^{2}+4(\gamma^{2}+q\beta^{2}+q^{2}\epsilon^{2})-4k_{\mu}q]z^{2\mu}+ [-2(\mu-1)(2\mu-1)qS_{0}- \nonumber \\ 4S_{0}(\beta^{2}+2q\epsilon^{2})+4k_{\mu}S_{0}]z^{\mu}+4S_{0}^{2}\epsilon^{2}+(\mu-1)^{2}S_{0}^{2}\big]^{\frac{1}{2}}\big\}.
\end{eqnarray}
For the requirement of $\pi_{f}(z)$ to be a $\mu th$ degree function, parameter $k_{\mu}$ which is given by $k=k_{\mu}z^{\mu -1}$ must be chosen properly;
\begin{equation}\label{kmu}
k_{\mu_{1,2}}=\frac{1}{2}[2\beta^{2}+\mu(\mu-1)q\pm \sqrt{(\mu^2 q^2+4\gamma^2)((\mu-1)^2+4\epsilon^2)}]
\end{equation}
Taking into account the $\pm$  signs in Eq.(\ref{kmu}), four different forms of $\pi_{f}(z)$ are obtained. The function $\pi_{f}(z)$ which is chosen as the function $\tau_{f}(z)$ given by  Eq.(\ref{eq:fh}) has a negative derivative for physical validity \cite{Nikiforov}.This condition is provided by;
\begin{equation}\label{kmu2}
k_{\mu}=\frac{1}{2}[2\beta^{2}+\mu(\mu-1)q - \sqrt{(\mu^2 q^2+4\gamma^2)((\mu-1)^2+4\epsilon^2)}].
\end{equation}
Using the chosen $k_{\mu}$ in Eq.(\ref{kmu2}) the function $\pi_{f}(z)$ is obtained as;
\begin{eqnarray}
\pi_{f}(z)=\frac{1}{2}\bigg[S_{0}\big(\mu-1+\sqrt{(\mu-1)^{2}+4\epsilon^{2}}\big)-\nonumber \\ \big(q(2\mu-1+\sqrt{(\mu -1)^{2}+4\epsilon^{2}})+\sqrt{\mu^{2}q^{2}+4\gamma^{2}}\big)z^{\mu}\bigg].
\end{eqnarray}
After determining $\pi_{f}(z)$, one can obtain the functions $\tau_{f}(z)$, $\lambda(z)$ and $\lambda_{n}(z)$ from Eq.(\ref{eq:fh}), Eq.(\ref{eq:fk}) and Eq.(\ref{eq:fp}) respectively;
\begin{equation}
\tau_{f}(z)=S_{0}\big(1+\sqrt{(\mu-1)^{2}+4\epsilon^{2}}\big)-\big(q(\mu+1+\sqrt{(\mu-1)^{2}+4\epsilon^{2}})+\sqrt{\mu^{2}q^{2}+4\gamma^{2}}\big)z^{\mu},
\end{equation}
\begin{eqnarray}
\lambda(z)=\frac{1}{2}\big[2\beta^{2}-\mu^{2}q-\sqrt{(\mu^{2}q^{2}+4\gamma^{2})((\mu-1)^{2}+4\epsilon^{2})}-\nonumber \\
\mu \sqrt{\mu^{2}q^{2}+4\gamma^{2}}-\mu q \sqrt{(\mu-1)^{2}+4\epsilon^{2}}\big]z^{\mu-1},
\end{eqnarray}
\begin{equation}
\lambda_{n}(z)=n\mu\big[q(\mu +1+\sqrt{(\mu-1)^{2}+4\epsilon^{2}})+\sqrt{\mu^{2}q^{2}+4\gamma^{2}}+\frac{(n-1)(\mu+1)q}{2}\big]z^{\mu-1}.
\end{equation}
For $\lambda (z)= \lambda_{n}(z)$, eigenvalue spectra of the problem is established by recalling the equalities given by Eq.(\ref{qntts});
\begin{eqnarray}
E^{2}-m^{2}=\frac{\alpha^{2}}{4}\bigg\{(\mu-1)^{2}- \nonumber \\
\big[\frac{4mS_{0}-\mu^{2}\alpha^{2}q-\mu \alpha\sqrt{\mu^{2}\alpha^{2}q^{2}+4S_{0}^{2}}(1+2n)-\mu(\mu+1)n(n+1)q\alpha^{2})}{\mu \alpha^{2}q(2n+1)+\alpha\sqrt{\mu^{2}\alpha^{2}q^{2}+4S_{0}^{2}}}\big]^{2}\bigg\}.
\end{eqnarray}
In order to obtain eigenfunction solution, the function $\phi(z)$ is determined by using Eq.(\ref{fi});
\begin{equation}\label{phir}
\phi(z)=z^{\frac{1}{2}(\mu-1+\sqrt{(\mu-1)^{2}+4\epsilon^{2}})}(S_{0}-qz^{\mu})^{\frac{1}{2\mu q}(\mu q+\sqrt{\mu^{2} q^{2}+4\gamma^{2}})}.
\end{equation}
Then, the functions $\rho(z)$ and $y_{n}(z)$ are obtained from Eq.(\ref{rho}) and Eq.(\ref{rod});
\begin{equation}
\rho(z)=z^{\sqrt{(\mu-1)^{2}+4\epsilon^{2}}}(S_{0}-qz^{\mu})^{\frac{1}{\mu q}(\sqrt{\mu^{2} q^{2}+4\gamma^{2}})}.
\end{equation}
\begin{eqnarray}\label{rodr}
y_{n}(z)=B_{n}z^{-\sqrt{(\mu-1)^{2}+4\epsilon^{2}}}(S_{0}-qz^{\mu})^{-\frac{1}{\mu q}(\sqrt{\mu^{2} q^{2}+4\gamma^{2}})}\nonumber \\
\frac{d^{n}}{dz^{n}}\big[z^{n+\sqrt{(\mu-1)^{2}+4\epsilon^{2}}}(S_{0}-qz^{\mu})^{n+\frac{1}{\mu q}(\sqrt{\mu^{2} q^{2}+4\gamma^{2}})}\big].
\end{eqnarray}
The right hand sides of Eq.(\ref{phir}) and Eq.(\ref{rodr}) are inserted in the transformation $\psi(z)=\phi(z)y_{n}(z)$:
\begin{eqnarray}
\psi(z)=B_{n}z^{\frac{1}{2}(\mu-1+\sqrt{(\mu-1)^{2}+4\epsilon^{2}})}(S_{0}-qz^{\mu})^{\frac{1}{2\mu q}(\mu q+\sqrt{\mu^{2} q^{2}+4\gamma^{2}})}\nonumber \\
z^{-\sqrt{(\mu-1)^{2}+4\epsilon^{2}}}(S_{0}-qz^{\mu})^{-\frac{1}{\mu q}(\sqrt{\mu^{2} q^{2}+4\gamma^{2}})}\nonumber \\
\frac{d^{n}}{dz^{n}}\big[z^{n+\sqrt{(\mu-1)^{2}+4\epsilon^{2}}}(S_{0}-qz^{\mu})^{n+\frac{1}{\mu q}(\sqrt{\mu^{2} q^{2}+4\gamma^{2}})}\big].
\end{eqnarray}
Consequently the eigenvalue and the eigenfunction spectra of a spinless particle in the generalized Hulthen potential which are identical to results in Ref.\cite{egrifes} and Ref.\cite{ikh} for $\mu=1$, have been obtained completely in view of conformable fractional calculus.
\section{Results and Discussion}
\begin{figure}
\centering
\includegraphics{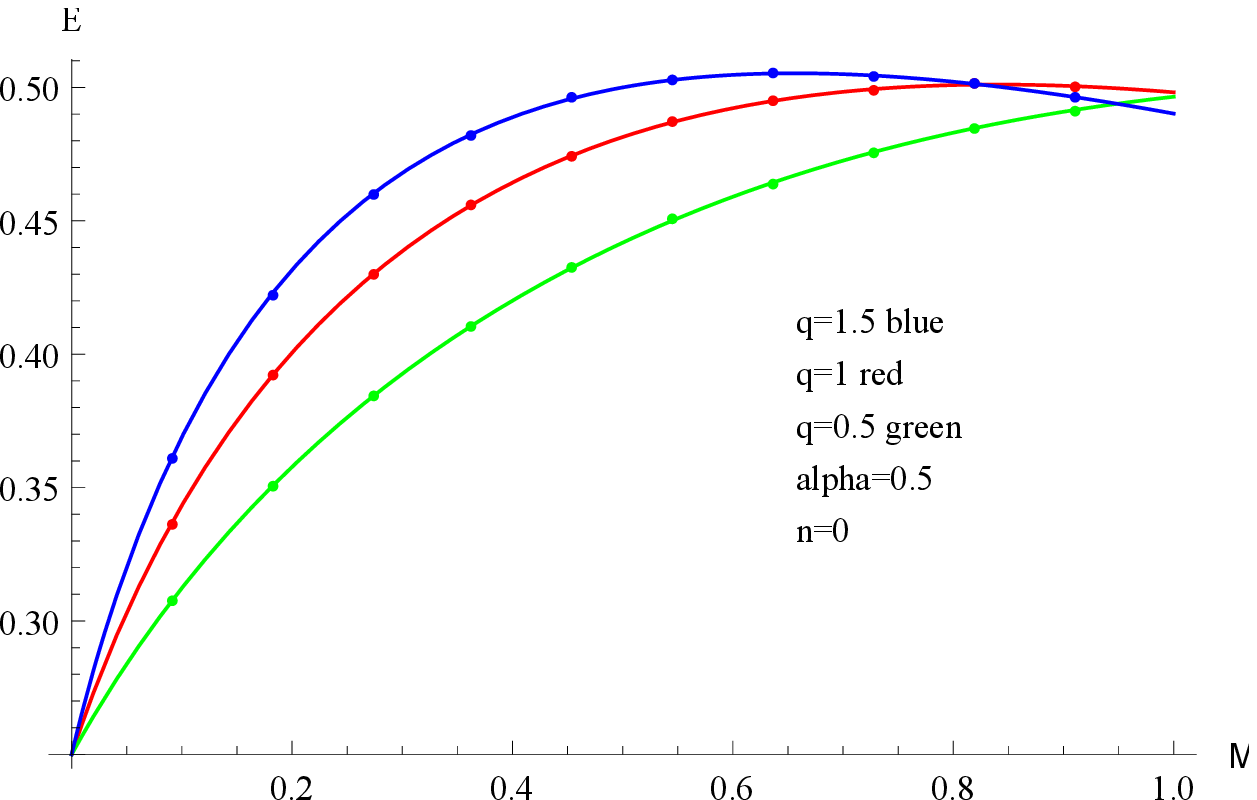}
\caption{The variation of the ground state energy of a relativistic particle moving in the generalized Hulthen potential as a function of the fractional order $\mu$ for three different values of potential deformation parameter $q$ where $\alpha=0.5$. }
\label{fig:1}       
\end{figure}

\begin{figure}

\centering
\includegraphics{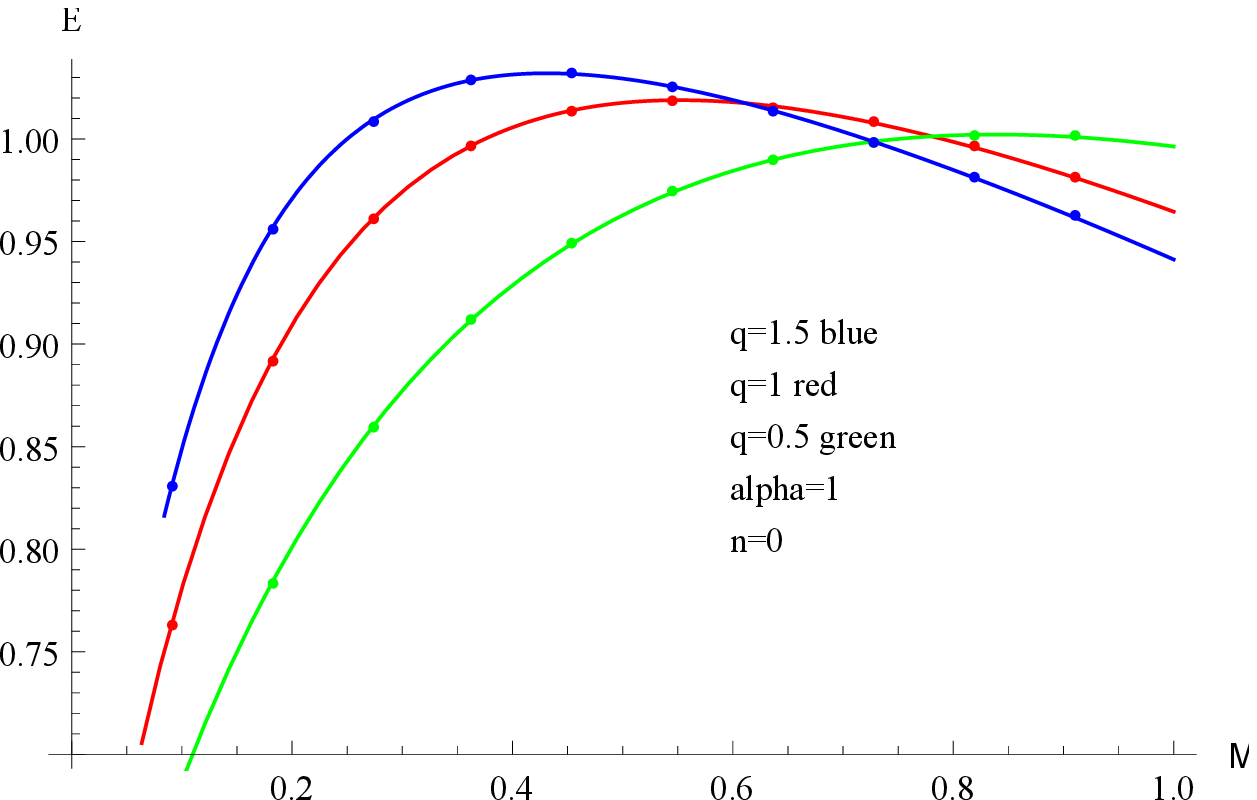}
\caption{The variation of the ground state energy of a relativistic particle moving in the generalized Hulthen potential as a function of the fractional order $\mu$ for three different values of potential deformation parameter $q$ where $\alpha=1$. }
\label{fig:2}       
\end{figure}

\begin{figure}
\centering
\includegraphics{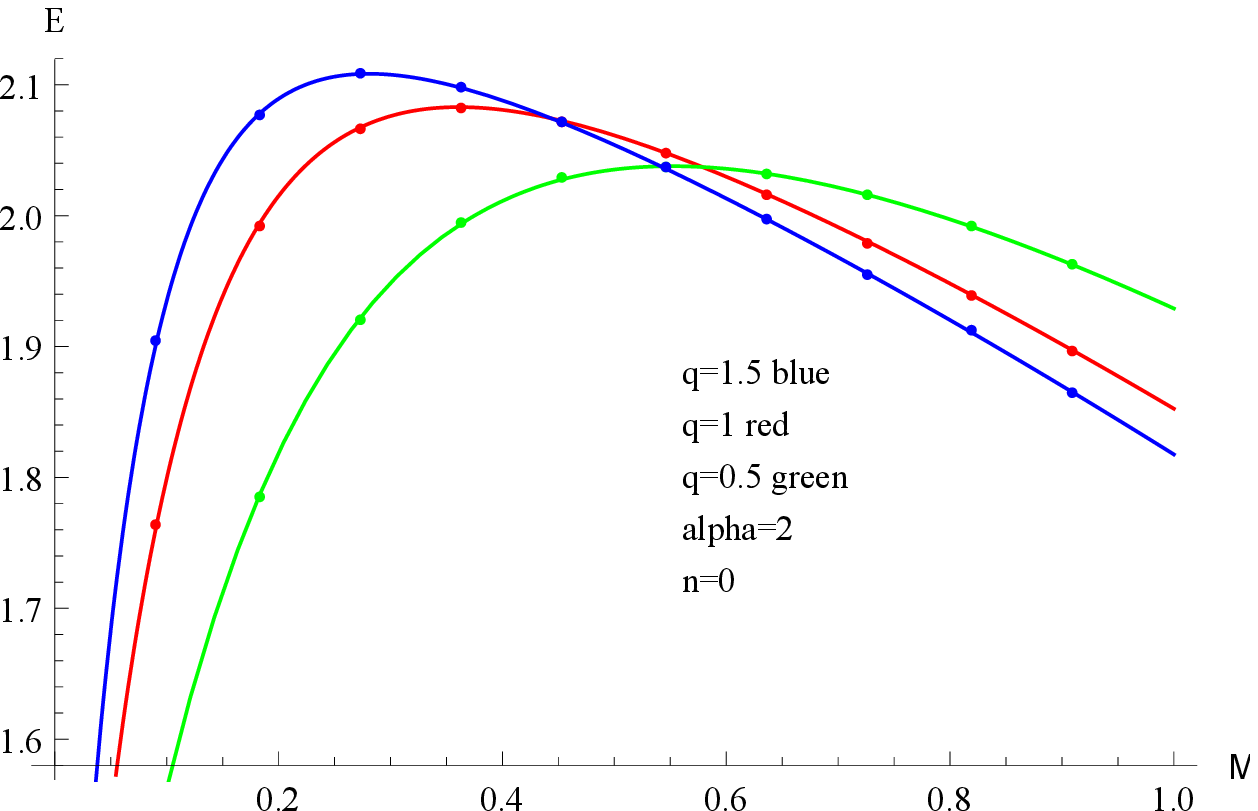}
\caption{The variation of the ground state energy of a relativistic particle moving in the generalized Hulthen potential as a function of the fractional order $\mu$ for three different values of potential deformation parameter $q$ where $\alpha=2$. }
\label{fig:3}       
\end{figure}

Fractionalization of the relativistic wave equations have been widely studied  by using Riemann-Liouville or Caputo fractional derivative operator in general. Since all fractional derivative operators have nonlocal character and they do not satisfy Leibniz rule, the wave equations including these operators are so complicated in order to obtain an analytical solution related to the fractional dimension of the space. Herein, a local fractional derivative operator is needed to arrive an exact solution. Local fractional form of the KG equation is proposed in order to describe the dynamics of a relativistic particle moving in the generalized Hulthen potential by means of conformable fractional derivative operator. Therefore variation of the energy and the wavefunction spectra with respect to the fractional order can be obtained in a more realistic manner. 
In the presented figures, evolution of ground state energy of a spinless particle in deformed Hulthen potential is represented as a function of the fractional order $\mu$ for three different values of the deformation parameter $q$ and for three different values of the range parameter $\alpha$, namely $0.5$, $1$ and $2$. It can be seen that the curves increase  more rapidly with increasing $q$ to a particular value of $\mu$. Then they decrease to the well known values of ground state energy at $\mu=1$ when green line in Figure~(\ref{fig:1}) is excluded.  
In Figure~(\ref{fig:2}) and Figure~(\ref{fig:3}) the initial values of the curves start at $\mu \neq 0$ for all values of $q$. Moreover maximum values of the curves are in evidence when $\mu$ reaches to the value $1$. On the whole, all curves intersect two by two at different points which correspond to the different values of $\mu$ and the curves reach maximum values more rapidly with increasing $\alpha$.
Moreover ground state energy for the deformed Hulthen potential is given numerically for fixed $S_{0}=0.25$ and given $\alpha$ in Table~(\ref{tab:1}), Table~(\ref{tab:2}) and Table~(\ref{tab:3}).


\clearpage

\clearpage

\clearpage

\clearpage
\begin{table}
\caption{Ground-state energy of the local fractional KG equation for $S_{0}=0.25$ and $\alpha=0.5$. Here $\alpha$ is expressed in units of Compton wavelenght, $\alpha=1/\lambda_{C}=mc/\hbar$}.
\label{tab:1}       
\centering
\begin{tabular}{lllll}
\hline\noalign{\smallskip}
 & $\mu=0.25$ &$\mu=0.5$ &$\mu=0.75$ & $\mu=1$ \\
$q$ & $E_{0}$ &$E_{0}$ & $E_{0}$ & $E_{0}$ \\
\noalign{\smallskip}\hline\noalign{\smallskip}
0.5& 0.376295 & 0.441808 & 0.478148 & 0.496505 \\
1& 0.42148 & 0.48417 & 0.5 & 0.498157 \\
1.5& 0.45227 &  0.5 & 0.503884 & 0.490179 \\
\noalign{\smallskip}\hline
\end{tabular}
\end{table}

\begin{table}
\caption{Ground-state energy of the local fractional KG equation for $S_{0}=0.25$ and $\alpha=1$. Here $\alpha$ is expressed in units of Compton wavelenght, $\alpha=1/\lambda_{C}=mc/\hbar$}.
\label{tab:2}       
\centering
\begin{tabular}{lllll}
\hline\noalign{\smallskip}
 & $\mu=0.25$ &$\mu=0.5$ &$\mu=0.75$ & $\mu=1$ \\
$q$ & $E_{0}$ &$E_{0}$ & $E_{0}$ & $E_{0}$ \\
\noalign{\smallskip}\hline\noalign{\smallskip}
0.5& 0.84296 & 0.962835 & 1 & 0.996314 \\
1& 0.947387 & 1.01761 & 1.00519 & 0.964541 \\
1.5& 1 & 1.02942 & 0.994548 & 0.941246 \\
\noalign{\smallskip}\hline
\end{tabular}
\end{table}

\begin{table}
\caption{Ground-state energy of the local fractional KG equation for $S_{0}=0.25$ and $\alpha=2$. Here $\alpha$ is expressed in units of Compton wavelenght, $\alpha=1/\lambda_{C}=mc/\hbar$}.
\label{tab:3}       
\centering
\begin{tabular}{lllll}
\hline\noalign{\smallskip}
 & $\mu=0.25$ &$\mu=0.5$ &$\mu=0.75$ & $\mu=1$ \\
$q$ & $E_{0}$ &$E_{0}$ & $E_{0}$ & $E_{0}$ \\
\noalign{\smallskip}\hline\noalign{\smallskip}
0.5& 1.89477 & 2.03522 & 2.01038 & 1.92908 \\
1& 2.05619 & 2.06136 & 1.97015 & 1.85251 \\
1.5& 2.1062 & 2.05407 & 1.94451 & 1.81759 \\
\noalign{\smallskip}\hline
\end{tabular}
\end{table}
\providecommand{\href}[2]{#2}

\clearpage

\address{
Department of Physics, Faculty of Science, Kirklareli University, Kirklareli, Turkey\\
\email{hale.karayer@gmail.com}\\
}

\address{
Faculty of Sport Sciences, Ege University,35100 Bornova, Izmir, Turkey\\
\email{dogan.demirhan@ege.edu.tr}\\
}

\address{
Department of Physics, Faculty of Science, Ege University,35100 Bornova, Izmir, Turkey\\
\email{fevzi.buyukkilic@ege.edu.tr}\\
}

\end{document}